\documentclass[
aps,%
12pt,%
final,%
notitlepage,%
oneside,%
onecolumn,%
nobibnotes,%
nofootinbib,%
superscriptaddress,%
noshowpacs]%
{revtex4}
\usepackage{graphicx}
\usepackage{color}
\usepackage[colorlinks]{hyperref}

\begin{document}

\bigskip\bigskip  \hrule\smallskip\hrule
\vspace{35mm}

\title{Testing General Relativity with Black Hole X-ray Data\footnote{Paper presented at the Fourth Zeldovich 
meeting, an international conference in honor of Ya. B. Zeldovich held in Minsk, Belarus on September 7--11, 2020.}}

\author{\bf \copyright $\:$  2021.
\quad \firstname{Cosimo}~\surname{Bambi}}%
\email{bambi@fudan.edu.cn}
\affiliation{Center for Field Theory and Particle Physics and Department of Physics, Fudan University, 200438 Shanghai, China}

\begin{abstract}

Einstein's theory of General Relativity was proposed over 100 years ago and has successfully passed a large number of observational tests in weak gravitational fields. On the contrary, the strong field regime is still largely unexplored. Astrophysical black holes are ideal laboratories for testing gravity in the strong field regime. Here I will briefly review the current efforts of my group to test the strong gravitational field around black holes using X-ray data from \textsl{NuSTAR}, \textsl{RXTE}, \textsl{Suzaku}, and \textsl{XMM-Newton}.
\end{abstract}

\maketitle

\section{Introduction}

The theory of General Relativity was proposed by Einstein at the end of 1915 and is still our standard framework for the description of gravitational fields and of the spacetime structure. The theory has been extensively tested in weak gravitational fields with experiments in the Solar System and observations of binary pulsars~\cite{Will:2014kxa}. The interest has now shifted to test General Relativity in other regimes. The past 20~years have seen significant efforts to try to verify the theory on very large scales with cosmological tests; this line of research is mainly motivated by the problems of dark matter and dark energy. Thanks to new observational facilities, in the past 10~years we have started testing General Relativity in the strong field regime with astrophysical compact objects.

The gravitational fields around black holes are the strongest gravitational fields that can be found today in the Universe, which makes these objects ideal laboratories for testing General Relativity in the strong field regime. Black holes can be tested with electromagnetic and gravitational wave techniques. Electromagnetic tests can investigate the interactions between the gravity and the matter sectors and can test the Einstein Equivalence Principle (motion of test-particles and non-gravitational physics in a gravitational field)~\cite{Bambi:2015kza,Krawczynski:2018fnw}. Gravitational wave tests are sensitive to the Einstein Equations~\cite{Berti:2018cxi,Berti:2018vdi}. The two methods are complementary and a comparison of their constraining power is only possible under specific conditions~\cite{Cardenas-Avendano:2016zml,Cardenas-Avendano:2019zxd}.

\begin{figure}
\includegraphics[width=0.7\textwidth]{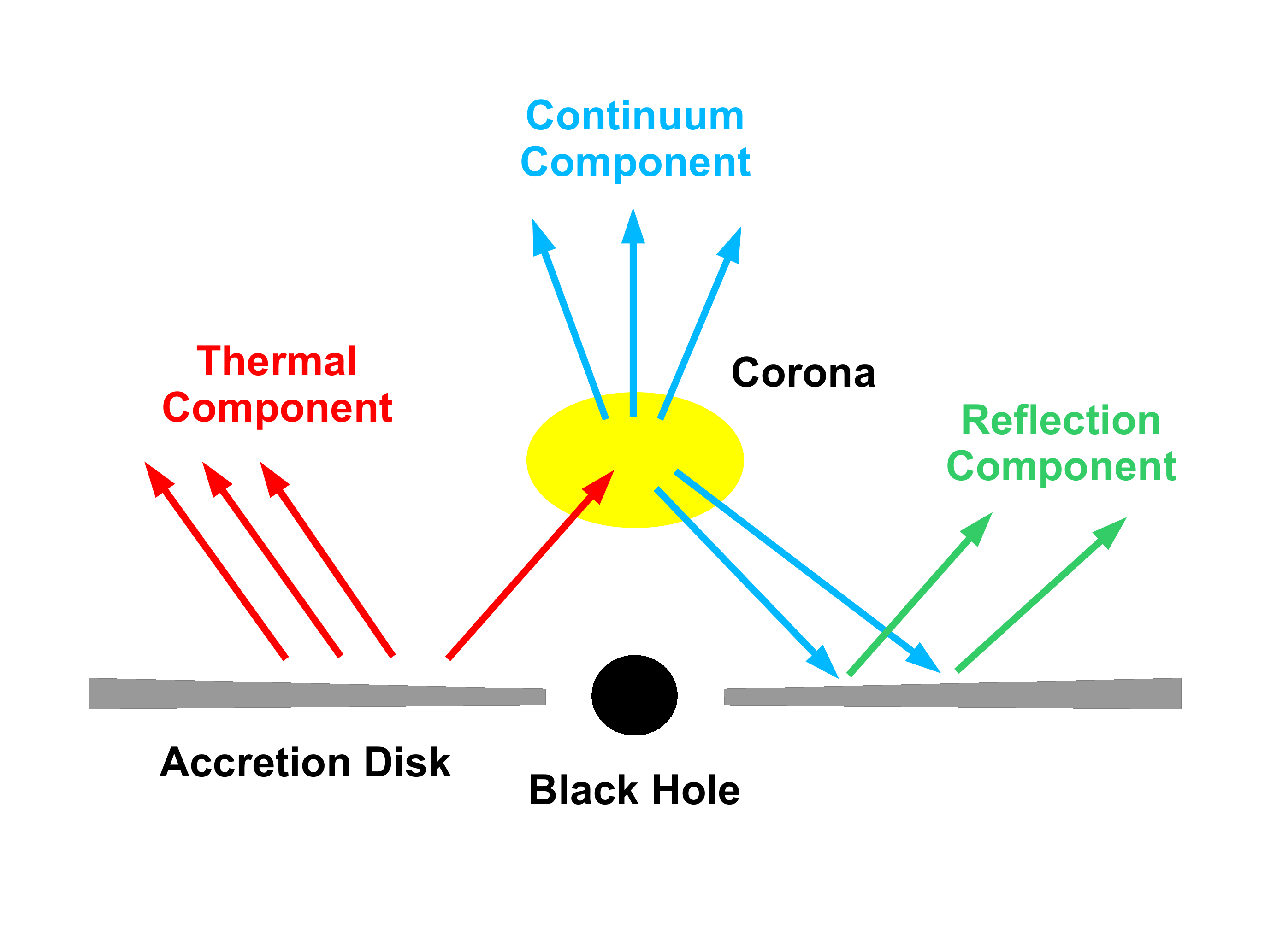}
\vspace{-1.0cm}
\caption{Black hole accreting from a geometrically thin and optically thick accretion disk and with a hot corona. \label{f-corona}}
\end{figure}

Here I will briefly review the efforts of my group and my collaborators to test the strong gravitational field of accreting black holes by the analysis of the X-ray radiation emitted from the inner part of their accretion disks. The astrophysical system of our studies is sketched in Fig.~\ref{f-corona}. The central black circle indicates the black hole, which is surrounded by a geometrically thin and optically thick accretion disk, in grey. Every point on the accretion disk is in local thermal equilibrium and has a blackbody-like spectrum. The whole disk has a multi-temperature blackbody-like spectrum: its emission is peaked in the soft X-ray band in the case of stellar-mass black holes and in the optical/UV band in the case of supermassive black holes. Thermal photons from the disk, indicated by red arrows in Fig.~\ref{f-corona}, can inverse Compton scatter off free electrons in the corona, the yellow region, which is some hotter ($\sim$100~keV) gas near the black hole and the inner part of the accretion disk. The Comptonized photons generate a continuum (blue arrows), which can be normally described by a power law component with a high energy exponential cut-off. A fraction of the Comptonized photons illuminate the accretion disk, producing a reflection component (green arrows). More details can be found in~\cite{Bambi:2017iyh} and references therein.


\section{X-ray reflection spectroscopy}

X-ray reflection spectroscopy refers to the analysis of the reflection features produced by illumination of a cold disk by a hot corona. In the past decade, this technique has been mainly developed for measuring black hole spins~\cite{Brenneman:2006hw,Reynolds:2019uxi}. The most prominent features in the reflection spectrum are usually the iron K$\alpha$ complex in the soft X-ray band and the Compton hump peaked at 20-30~keV. The reflection spectrum of the accretion disk detected by a distant observer is the result of the reflection spectrum at the emission point and relativistic effects (Doppler boosting, gravitational redshift, and light bending) occurring in the strong gravity region around the compact object. In the presence of the correct astrophysical model and high-quality data, the analysis of these reflection features can be used to test fundamental physics.

Extending early work in literature calculating iron line profiles from accretion disks in non-Kerr spacetimes~\cite{Schee:2008fc,Johannsen:2012ng,Bambi:2012at}, we are developing the {\tt XSPEC}-compatible relativistic reflection model {\tt relxill\_nk}~\cite{Bambi:2016sac,Abdikamalov:2019yrr}\footnote{A public version is available at\\\url{http://www.physics.fudan.edu.cn/tps/people/bambi/Site/RELXILL_NK.html}.}, which is an extension of the {\tt relxill} package~\cite{Dauser:2013xv,Garcia:2013lxa} to non-Kerr spacetimes. The model has been mainly used with the Johannsen metric~\cite{Johannsen:2015pca}, which is a parametric black hole metric in which deviations from the Kerr solution are quantified by an infinite number of deformation parameters. Employing {\tt relxill\_nk} to fit the relativistic reflection component in black hole X-ray data, we can measure the model parameters, including the deformation parameters. As in a null-experiment, we can verify if observations are consistent with vanishing deformation parameters and thus if they can confirm the Kerr spacetime around black holes. The model can be used even to test specific non-Kerr black hole solutions in theories beyond General Relativity~\cite{Zhou:2018bxk,Zhou:2019hqk,Zhu:2020cfn}.

Tab.~\ref{t-rxnk1} and Tab.~\ref{t-rxnk2} show our current published measurements of the deformation parameter $\alpha_{13}$ of the Johannsen metric (assuming that all other deformation parameters vanish) from the analysis of X-ray data of, respectively, stellar-mass black holes in X-ray binaries and supermassive black holes in AGNs. The Kerr metric corresponds to $\alpha_{13} = 0$. More details can be found in the original articles~\cite{Liu:2019vqh,Zhang:2019ldz,Abdikamalov:2020oci,Xu:2018lom,Wang-Ji:2018ssh,Tripathi:2019bya,Cao:2017kdq,Bambi:2018ggp,Tripathi:2018bbu,Liu:2020fpv,Tripathi:2018lhx,Choudhury:2018zmf}.

As shown in Tab.~\ref{t-rxnk1}, in Ref.~\cite{Liu:2019vqh} we were not able to constrain $\alpha_{13}$ in Cygnus~X-1. This is mainly due to the complexity of the source: Cygnus~X-1 is a high-mass X-ray binary and its spectrum is significantly affected by the strong wind from the companion star, which challenges a robust test of the spacetime metric. In Tab.~\ref{t-rxnk2}, two measurements are not consistent with the Kerr metric at 90\% of confidence level. In the case Fairall~9, the Kerr metric is recovered at 2-$\sigma$. For RBS~1124, the fit provides a few measurements because $\chi^2$ has multiple minima; Tab.~\ref{t-rxnk2} just shows the measurement with the lowest $\chi^2$, but actually another measurement with a slightly higher $\chi^2$ is consistent with the Kerr geometry. Note that, even if precise and consistent with the Kerr solution, the measurements of $\alpha_{13}$ from \textsl{Suzaku} data of AGNs derived in \cite{Tripathi:2019bya} should be taken with some caution, as they are based only on the analysis of the soft X-ray band (0.6-10~keV) and some sources are thought to accrete above the thin disk limit, where {\tt relxill\_nk} should not be used~\cite{Riaz:2019bkv,Riaz:2019kat}.


\section{Continuum-fitting method}

In the Kerr spacetime, the multi-temperature blackbody spectrum of the accretion disk as detected by a distant observer depends on five parameters: the black hole mass $M$, the black hole distance $D$, the viewing angle of the disk $i$, the mass accretion rate $\dot{M}$, and the black hole spin parameter $a_*$. It is not possible to fit the data of a source and measure the values of these five parameters because there is a parameter degeneracy. However, if we can get independent measurements of $M$, $D$, and $i$, for example from optical observations, we can fit the data and infer the values of $a_*$ and $\dot{M}$. This is the continuum-fitting method~\cite{Zhang:1997dy,McClintock:2013vwa}.

As in the case of X-ray reflection spectroscopy, even the continuum-fitting method can be extended to non-Kerr spacetimes and be used to test the Kerr metric around black holes~\cite{Bambi:2011jq,Bambi:2012tg}. The technique can be normally applied only to stellar-mass black holes in X-ray binaries. In the case of supermassive black holes, the thermal component mainly emits in the optical/UV band, and extinction limits the ability to get accurate measurements of the spectrum.

Our {\tt XSPEC} model to calculate thermal spectra of thin disks in non-Kerr spacetimes is called {\tt nkbb}~\cite{Zhou:2019fcg}. For the moment, we have only applied the model to measure the deformation parameter $\alpha_{13}$ of the Johannsen metric from \textsl{RXTE} data of the stellar-mass black hole in LMC~X-1 in Ref.~\cite{Tripathi:2020qco}. Our measurement is shown in Tab.~\ref{t-nkbb}. Even in the presence of high-quality data, the thermal spectrum is often degenerate with respect to the black hole spin and the deformation parameter, so this technique can normally provide weak constraints on possible deviations from the Kerr solution with respect to the analysis of the relativistic reflection features. However, it is potentially possible to test a source with both techniques, and this can help to get stronger and more robust measurements of the deformation parameters.


\section{Concluding remarks}

Here I have very briefly reviewed the state of the art in tests of the Kerr nature of astrophysical black holes using X-ray data. This is quite a new field and its techniques can be extended to test other predictions of General Relativity, in particular new interactions between the gravity and the matter sectors. Note that our current constraints on the Johannsen deformation parameter $\alpha_{13}$ obtained with {\tt relxill\_nk} are an order of magnitude more stringent than the constraint that can be obtained from the available data of the Event Horizon Telescope~\cite{Psaltis:2020lvx}.

We are now improving the astrophysical models to study and limit the systematic uncertainties in the measurements of the deformation parameters. Work is in progress to test a stellar-mass black hole with both the analysis of its reflection features and the continuum-fitting method. At the moment, the exact mechanism responsible for the quasi-periodic oscillations (QPOs) in the X-ray spectra of black holes is not well understood. However, for the future we can expect that even the measurement of the QPO frequencies of a source can test the strong gravitational field around a black hole. In such a case, we will have the possibility of testing the same object with three different methods, and presumably get more stringent and robust constraints on possible deviations from the predictions of Einstein's theory of General Relativity.


\section*{Funding}

This work was supported by the Innovation Program of the Shanghai Municipal Education Commission, Grant No.~2019-01-07-00-07-E00035, the National Natural Science Foundation of China (NSFC), Grant No.~11973019, and Fudan University, Grant No.~JIH1512604.


\clearpage

\begin{table}[!h]
\caption{Constraints on the Johannsen deformation parameter $\alpha_{13}$ from stellar-mass black holes using {\tt relxill\_nk}. The reported uncertainties correspond to the 90\% CL ($\Delta\chi^2 = 2.71$). \label{t-rxnk1}}
\bigskip
\begin{tabular}{cccc}
\hline\hline
Source & \hspace{0.8cm} $\alpha_{13}$ \hspace{0.8cm} & \hspace{0.5cm} Instrument(s) \hspace{0.5cm} & \hspace{0.5cm} Reference(s) \hspace{0.5cm} \\
\hline
Cygnus~X-1 & unconstrained & \textsl{NuSTAR} & \cite{Liu:2019vqh} \\
\hline
GRS~1915+105 & $0.00_{-0.15}^{+0.05}$ & \textsl{Suzaku} & \cite{Zhang:2019ldz,Abdikamalov:2020oci} \\
\hline
GS~1354--645 & $0.06_{-0.31}^{+0.05}$ & \textsl{NuSTAR} & \cite{Xu:2018lom} \\
\hline
GX~339--4 & $-0.8_{-0.6}^{+0.8}$ & \textsl{RXTE} & \cite{Wang-Ji:2018ssh} \\
\hline\hline
\end{tabular}
\end{table}

\begin{table}[!h]
\caption{Constraints on the Johannsen deformation parameter $\alpha_{13}$ from supermassive black holes using {\tt relxill\_nk}. The reported uncertainties correspond to the 90\% CL ($\Delta\chi^2 = 2.71$). \label{t-rxnk2}}
\bigskip
\begin{tabular}{cccc}
\hline\hline
Source & \hspace{0.8cm} $\alpha_{13}$ \hspace{0.8cm} & \hspace{0.5cm} Instrument(s) \hspace{0.5cm} & \hspace{0.5cm} Reference(s) \hspace{0.5cm} \\
\hline
1H0419--577 & $0.00_{-0.14}^{+0.04}$ & \textsl{Suzaku} & \cite{Tripathi:2019bya} \\
\hline
1H0707--495 & $-2 \sim 1$ & \textsl{XMM-Newton} & \cite{Cao:2017kdq,Bambi:2018ggp} \\
\hline
Ark~120 & $0.00_{-0.08}^{+0.01}$ & \textsl{Suzaku} & \cite{Tripathi:2019bya} \\
\hline
Ark~564 & $-1.0_{-0.2}^{+1.0}$ & \textsl{Suzaku} & \cite{Tripathi:2018bbu} \\
\hline
Fairall~9 & $< -0.3$ & \textsl{XMM-Newton}+\textsl{NuSTAR} & \cite{Liu:2020fpv} \\
\hline
MCG--6--30--15 & $0.00_{-0.20}^{+0.07}$ & \textsl{XMM-Newton}+\textsl{NuSTAR} & \cite{Tripathi:2018lhx} \\
\hline
Mrk~335 & $0.2_{-1.1}^{+0.1}$ & \textsl{Suzaku} & \cite{Choudhury:2018zmf} \\
\hline
PKS~0558--504 & $0.03_{-0.20}^{+0.02}$ & \textsl{Suzaku} & \cite{Tripathi:2019bya} \\
\hline
RBS~1124 & $-0.7_{-0.1}^{+0.1}$ & \textsl{Suzaku} & \cite{Tripathi:2019bya} \\
\hline
Swift~J0501.9--3239 & $0.00_{-0.07}^{+0.03}$ & \textsl{Suzaku} & \cite{Tripathi:2019bya} \\
\hline
Ton~S180 & $0.01_{-0.32}^{+0.02}$ & \textsl{Suzaku} & \cite{Tripathi:2019bya} \\
\hline\hline
\end{tabular}
\end{table}

\begin{table}[!h]
\caption{Constraints on the Johannsen deformation parameter $\alpha_{13}$ from stellar-mass black holes using {\tt nkbb}. The reported uncertainties correspond to the 90\% CL ($\Delta\chi^2 = 2.71$). \label{t-nkbb}}
\bigskip
\begin{tabular}{cccc}
\hline\hline
\hspace{0.5cm} Source \hspace{0.5cm} & \hspace{0.5cm} $\alpha_{13}$ \hspace{0.5cm} & \hspace{0.5cm} Instrument(s) \hspace{0.5cm} & \hspace{0.5cm} Reference(s) \hspace{0.5cm} \\
\hline
LMC~X-1 & $0.32_{-3.1}^{+0.04}$ & \textsl{RXTE} & \cite{Tripathi:2020qco} \\
\hline\hline
\end{tabular}
\end{table}


\clearpage



\begin{thebibliography}{99}

\bibitem{Will:2014kxa}
C.~M.~Will,
Living Rev. Rel. \textbf{17}, 4 (2014)
[arXiv:1403.7377 [gr-qc]].

\bibitem{Bambi:2015kza}
C.~Bambi,
Rev. Mod. Phys. \textbf{89}, 025001 (2017)
[arXiv:1509.03884 [gr-qc]].

\bibitem{Krawczynski:2018fnw}
H.~Krawczynski,
Gen. Rel. Grav. \textbf{50}, 100 (2018)
[arXiv:1806.10347 [astro-ph.HE]].

\bibitem{Berti:2018cxi}
E.~Berti, K.~Yagi and N.~Yunes,
Gen. Rel. Grav. \textbf{50}, 46 (2018)
[arXiv:1801.03208 [gr-qc]].

\bibitem{Berti:2018vdi}
E.~Berti, K.~Yagi, H.~Yang and N.~Yunes,
Gen. Rel. Grav. \textbf{50}, 49 (2018)
[arXiv:1801.03587 [gr-qc]].

\bibitem{Cardenas-Avendano:2016zml}
A.~Cardenas-Avendano, J.~Jiang and C.~Bambi,
Phys. Lett. B \textbf{760}, 254-258 (2016)
[arXiv:1603.04720 [gr-qc]].

\bibitem{Cardenas-Avendano:2019zxd}
A.~Cardenas-Avendano, S.~Nampalliwar and N.~Yunes,
Class. Quant. Grav. \textbf{37}, 135008 (2020)
[arXiv:1912.08062 [gr-qc]].

\bibitem{Bambi:2017iyh}
C.~Bambi,
Annalen Phys. \textbf{530}, 1700430 (2018)
[arXiv:1711.10256 [gr-qc]].

\bibitem{Brenneman:2006hw}
L.~W.~Brenneman and C.~S.~Reynolds,
Astrophys. J. \textbf{652}, 1028-1043 (2006)
[arXiv:astro-ph/0608502 [astro-ph]].

\bibitem{Reynolds:2019uxi}
C.~S.~Reynolds,
Nature Astron. \textbf{3}, 41-47 (2019)
[arXiv:1903.11704 [astro-ph.HE]].

\bibitem{Schee:2008fc}
J.~Schee and Z.~Stuchlik,
Gen. Rel. Grav. \textbf{41}, 1795-1818 (2009)
[arXiv:0812.3017 [astro-ph]].

\bibitem{Johannsen:2012ng}
T.~Johannsen and D.~Psaltis,
Astrophys. J. \textbf{773}, 57 (2013)
[arXiv:1202.6069 [astro-ph.HE]].

\bibitem{Bambi:2012at}
C.~Bambi,
Phys. Rev. D \textbf{87}, 023007 (2013)
[arXiv:1211.2513 [gr-qc]].

\bibitem{Bambi:2016sac}
C.~Bambi, A.~Cardenas-Avendano, T.~Dauser, J.~A.~Garcia and S.~Nampalliwar,
Astrophys. J. \textbf{842}, 76 (2017)
[arXiv:1607.00596 [gr-qc]].

\bibitem{Abdikamalov:2019yrr}
A.~B.~Abdikamalov, D.~Ayzenberg, C.~Bambi, T.~Dauser, J.~A.~Garcia and S.~Nampalliwar,
Astrophys. J. \textbf{878}, 91 (2019)
[arXiv:1902.09665 [gr-qc]].

\bibitem{Dauser:2013xv}
T.~Dauser, J.~Garcia, J.~Wilms, M.~Bock, L.~W.~Brenneman, M.~Falanga, K.~Fukumura and C.~S.~Reynolds,
Mon. Not. Roy. Astron. Soc. \textbf{430}, 1694 (2013)
[arXiv:1301.4922 [astro-ph.HE]].

\bibitem{Garcia:2013lxa}
J.~Garcia, T.~Dauser, A.~Lohfink, T.~R.~Kallman, J.~Steiner, J.~E.~McClintock, L.~Brenneman, J.~Wilms, W.~Eikmann, C.~S.~Reynolds and F.~Tombesi,
Astrophys. J. \textbf{782}, 76 (2014)
[arXiv:1312.3231 [astro-ph.HE]].

\bibitem{Johannsen:2015pca}
T.~Johannsen,
Phys. Rev. D \textbf{88}, 044002 (2013)
[arXiv:1501.02809 [gr-qc]].

\bibitem{Zhou:2018bxk}
M.~Zhou, Z.~Cao, A.~Abdikamalov, D.~Ayzenberg, C.~Bambi, L.~Modesto and S.~Nampalliwar,
Phys. Rev. D \textbf{98}, 024007 (2018)
[arXiv:1803.07849 [gr-qc]].

\bibitem{Zhou:2019hqk}
M.~Zhou, A.~B.~Abdikamalov, D.~Ayzenberg, C.~Bambi, L.~Modesto, S.~Nampalliwar and Y.~Xu,
Europhys. Lett. \textbf{125}, 30002 (2019)
[arXiv:2003.03738 [gr-qc]].

\bibitem{Zhu:2020cfn}
J.~Zhu, A.~B.~Abdikamalov, D.~Ayzenberg, M.~Azreg-Ainou, C.~Bambi, M.~Jamil, S.~Nampalliwar, A.~Tripathi and M.~Zhou,
Eur. Phys. J. C \textbf{80}, 622 (2020)
[arXiv:2005.00184 [gr-qc]].

\bibitem{Liu:2019vqh}
H.~Liu, A.~B.~Abdikamalov, D.~Ayzenberg, C.~Bambi, T.~Dauser, J.~A.~Garcia and S.~Nampalliwar,
Phys. Rev. D \textbf{99}, 123007 (2019)
[arXiv:1904.08027 [gr-qc]].

\bibitem{Zhang:2019ldz}
Y.~Zhang, A.~B.~Abdikamalov, D.~Ayzenberg, C.~Bambi and S.~Nampalliwar,
Astrophys. J. \textbf{884}, 147 (2019)
[arXiv:1907.03084 [gr-qc]].

\bibitem{Abdikamalov:2020oci}
A.~B.~Abdikamalov, D.~Ayzenberg, C.~Bambi, T.~Dauser, J.~A.~Garcia, S.~Nampalliwar, A.~Tripathi and M.~Zhou,
Astrophys. J. \textbf{899}, 80 (2020)
[arXiv:2003.09663 [astro-ph.HE]].

\bibitem{Xu:2018lom}
Y.~Xu, S.~Nampalliwar, A.~B.~Abdikamalov, D.~Ayzenberg, C.~Bambi, T.~Dauser, J.~A.~Garcia and J.~Jiang,
Astrophys. J. \textbf{865}, 134 (2018)
[arXiv:1807.10243 [gr-qc]].

\bibitem{Wang-Ji:2018ssh}
J.~Wang, A.~B.~Abdikamalov, D.~Ayzenberg, C.~Bambi, T.~Dauser, J.~A.~Garcia, S.~Nampalliwar and J.~F.~Steiner,
JCAP \textbf{05}, 026 (2020)
[arXiv:1806.00126 [gr-qc]].

\bibitem{Tripathi:2019bya}
A.~Tripathi, J.~Yan, Y.~Yang, Y.~Yan, M.~Garnham, Y.~Yao, S.~Li, Z.~Ding, A.~B.~Abdikamalov, D.~Ayzenberg, C.~Bambi, T.~Dauser, J.~A.~Garcia, J.~Jiang and S.~Nampalliwar,
Astrophys. J. \textbf{874}, 135 (2019)
[arXiv:1901.03064 [gr-qc]].

\bibitem{Cao:2017kdq}
Z.~Cao, S.~Nampalliwar, C.~Bambi, T.~Dauser and J.~A.~Garcia,
Phys. Rev. Lett. \textbf{120}, 051101 (2018)
[arXiv:1709.00219 [gr-qc]].

\bibitem{Bambi:2018ggp}
C.~Bambi, A.~B.~Abdikamalov, D.~Ayzenberg, Z.~Cao, H.~Liu, S.~Nampalliwar, A.~Tripathi, J.~Wang-Ji and Y.~Xu,
Universe \textbf{4}, 79 (2018)
[arXiv:1806.02141 [gr-qc]].

\bibitem{Tripathi:2018bbu}
A.~Tripathi, S.~Nampalliwar, A.~B.~Abdikamalov, D.~Ayzenberg, J.~Jiang and C.~Bambi,
Phys. Rev. D \textbf{98}, 023018 (2018)
[arXiv:1804.10380 [gr-qc]].

\bibitem{Liu:2020fpv}
H.~Liu, H.~Wang, A.~B.~Abdikamalov, D.~Ayzenberg and C.~Bambi,
Astrophys. J. \textbf{896}, 160 (2020)
[arXiv:2004.11542 [gr-qc]].

\bibitem{Tripathi:2018lhx}
A.~Tripathi, S.~Nampalliwar, A.~B.~Abdikamalov, D.~Ayzenberg, C.~Bambi, T.~Dauser, J.~A.~Garcia and A.~Marinucci,
Astrophys. J. \textbf{875}, 56 (2019)
[arXiv:1811.08148 [gr-qc]].

\bibitem{Choudhury:2018zmf}
K.~Choudhury, S.~Nampalliwar, A.~B.~Abdikamalov, D.~Ayzenberg, C.~Bambi, T.~Dauser and J.~A.~Garcia,
Astrophys. J. \textbf{879}, 80 (2019)
[arXiv:1809.06669 [gr-qc]].

\bibitem{Riaz:2019bkv}
S.~Riaz, D.~Ayzenberg, C.~Bambi and S.~Nampalliwar,
Mon. Not. Roy. Astron. Soc. \textbf{491}, 417-426 (2020)
[arXiv:1908.04969 [astro-ph.HE]].

\bibitem{Riaz:2019kat}
S.~Riaz, D.~Ayzenberg, C.~Bambi and S.~Nampalliwar,
Astrophys. J. \textbf{895}, 61 (2020)
[arXiv:1911.06605 [astro-ph.HE]].

\bibitem{Zhang:1997dy}
S.~N.~Zhang, W.~Cui and W.~Chen,
Astrophys. J. Lett. \textbf{482}, L155 (1997)
[arXiv:astro-ph/9704072 [astro-ph]].

\bibitem{McClintock:2013vwa}
J.~E.~McClintock, R.~Narayan and J.~F.~Steiner,
Space Sci. Rev. \textbf{183}, 295-322 (2014)
[arXiv:1303.1583 [astro-ph.HE]].

\bibitem{Bambi:2011jq}
C.~Bambi and E.~Barausse,
Astrophys. J. \textbf{731}, 121 (2011)
[arXiv:1012.2007 [gr-qc]].

\bibitem{Bambi:2012tg}
C.~Bambi,
Astrophys. J. \textbf{761}, 174 (2012)
[arXiv:1210.5679 [gr-qc]].

\bibitem{Zhou:2019fcg}
M.~Zhou, A.~B.~Abdikamalov, D.~Ayzenberg, C.~Bambi, H.~Liu and S.~Nampalliwar,
Phys. Rev. D \textbf{99}, 104031 (2019)
[arXiv:1903.09782 [gr-qc]].

\bibitem{Tripathi:2020qco}
A.~Tripathi, M.~Zhou, A.~B.~Abdikamalov, D.~Ayzenberg, C.~Bambi, L.~Gou, V.~Grinberg, H.~Liu and J.~F.~Steiner,
Astrophys. J. \textbf{897}, 84 (2020)
[arXiv:2001.08391 [gr-qc]].

\bibitem{Psaltis:2020lvx}
D.~Psaltis \textit{et al.} [EHT],
Phys. Rev. Lett. \textbf{125}, 141104 (2020)
[arXiv:2010.01055 [gr-qc]].

\end{thebibliography}
\end{document}